\documentclass[12pt]{article}   
\usepackage{latexsym} 
\usepackage{mathptm}
\usepackage{amsmath}
\usepackage{amssymb}
\usepackage{graphicx}

\def\beq{\begin{equation}}
\def\eeq{\end{equation}}
\def\beqn{\begin{eqnarray}}
\def\eeqn{\end{eqnarray}}

\begin{document}
 
\title{Comment on arXiv:1007.0718 by Lee 
Smolin}
\author{Sabine Hossenfelder \thanks{hossi@nordita.org}\\
{\footnotesize{\sl NORDITA, Roslagstullsbacken 23, 106 91 Stockholm, Sweden}}}
\date{}
\maketitle

\vspace*{-1cm}
\begin{abstract}
In a recent paper it was suggested a novel interpretation of deformed special
relativity. In that new approach, nonlocal effects that had previously been
shown to occur and be incompatible with experiment to high precision, are 
interpreted as coordinate artifacts that do not lead to real physical consequences.
It is argued here that if one follows through the consequences of this thought, one finds 
that the theory one is dealing with needs to be ordinary special relativity to
precision even better than the bound on nonlocal effect already requires.
Consequently, the new approach cannot be understood as a version of deformed special
relativity that circumvents the bound. 
\end{abstract}

\section{Introduction}

Deformations of special relativity ({\sc DSR}) \cite{AmelinoCamelia:2000mn,DSR,ReviewGAC,ReviewJKG} allegedly make it
possible to introduce an energy-dependent speed of light in position
space while still preserving observer-independence. In \cite{Hossenfelder:2009mu,Hossenfelder:2010tm} it
has been shown that such an energy-dependent and observer-independent speed
of light is in conflict with
observations, at least to first order in energy over Planck mass, $E/m_{\rm p}$. The reason
for this conflict is that the model causes macroscopic violations of locality that
are incompatible with well-confirmed particle interaction processes of the standard model. 

It was then claimed by Smolin in \cite{Smolin:2010xa} that quantum uncertainty may be
a possibility to hide the nonlocality despite the fact that the contrary was shown in \cite{Hossenfelder:2009mu,Hossenfelder:2010tm}. 
In \cite{Hossenfelder:2010yp} it was demonstrated that Smolin's ansatz is inconsistent and, 
when corrected, just reproduces the problem\footnote{Note that the reply \cite{Hossenfelder:2010yp} referred
to the first version of \cite{Smolin:2010xa}, which has since been significantly revised.}. In \cite{Jacob:2010vr}, Jacob et al found the same
nonlocality as \cite{Hossenfelder:2009mu,Hossenfelder:2010tm} and correctly
identified it as a glaring inconsistency, unfortunately without acknowledging they were
just reproducing the earlier result from \cite{Hossenfelder:2009mu,Hossenfelder:2010tm}. In \cite{AmelinoCamelia:2010qv} Amelino-Camelia et
al then shifted the nonlocality from the detector (Earth) to the source (gamma ray burst) and
thus moved the original setup by some Gpc. This is a possibility
that was already commented on in \cite{Hossenfelder:2009mu}. Moving the nonlocality does
of course not remove it. It was explained in \cite{Hossenfelder:2010jn} that Amelino-Camelia
et al's claim that moving the nonlocality changes the bound by more than 20 orders of
magnitude is due to them inaccurately using the duration of a gamma ray burst (of the order
seconds) for a single elementary particle collision (of the order fm/$c$). The location at
which the nonlocality becomes non-negligible depends on the exact scenario one 
considers as will become clear in the following. The bound itself however does not
significantly depend on where the non-locality becomes noticeable\footnote{It does slightly depend
on the location because one could argue that experimental tests of elementary particle
collisions in distant astrophysical sources are not as precise as those done on
Earth. However, we have so far no evidence that quantum field theory has worked any
different during the evolution of the universe than it does today on our planet. Thus, if these bounds are weaker after
shifting the nonlocality elsewhere, they
cannot be much weaker.}.

Now Smolin has put forward a new preprint \cite{Smolin:2010mx} with another attempted circumvention of 
the problem of nonlocality in {\sc DSR}. The ansatz he presents overlaps partly with the one of 
Amelino-Camelia et al in \cite{AmelinoCamelia:2010qv}, but it differs in the interpretation 
and adds some interesting new layers to the argument that I will discuss here. 

In the following we use units in which $c=\hbar=1$.

\section{Brief summary of the problem with nonlocality in DSR}

Let us start with explaining briefly how the bound in \cite{Hossenfelder:2009mu,Hossenfelder:2010tm}
was derived and introduce some terminology. Consider a source in the far distance (in the original
example, a gamma ray burst) that emits two photons. In {\sc DSR} the speed of the photon $c(E)$
depends on its energy such that deviations from the speed that we have measured 
become larger with larger energy. One photon is assumed to have arbitrarily low energy and
thus moves with the ordinary speed of light. This photon is only there as a reference and
represents the special relativistic limit. The other 
photon has a higher energy and is slowed down due to the energy-dependence of the speed of light\footnote{It could
also speed up, depending on the function $c(E)$. The following arguments however do not depend on
whether the speed decreases or increases with energy. We will thus stick with one case, that in which
the speed decreases, to improve readability.}. It travels a long distance till it collides with 
an electron in a detector (in the original example, on Earth).  The scattering process causes a macroscopic event. 
The delay between the arrival time
of the low and the high energetic photon is for a first order modification of the
speed of light
\beqn
\Delta T = \alpha \frac{E_\gamma}{m_{\rm p}} L \quad, \label{dt}
\eeqn
where $E_\gamma$ is the energy of the highly energetic photon, $m_{\rm p}$ is
the Planck scale, $L$ is the distance to the source, and $\alpha$ is a dimensionless
parameter of the model characterizing the strength of the effect. The above formula has to be corrected 
when one takes into account the cosmological expansion \cite{Ellis:2002in,Jacob:2008bw} but this will not matter 
for the following. The numbers we will use in the following for these quantities are $E_\gamma \approx 10$~GeV
and $L \approx 4$~Gpc, such that $\Delta T \approx \alpha 1$~s.

The question is now, how does the same process look like for an observer who
moves relative to the first restframe? To avoid having to carry around too many
different parameters, the energy of the electron has been assumed to me much
smaller than that of the highly energetic photon and the detector represents
another macroscopic and special relativistic limit. Thus, we know how the
worldlines of the electron and the detector have to be transformed. It's just
a normal Lorentz transformation. 

To transform the worldline of the highly energetic
photon, we cannot use the normal Lorentz transformation. We can however derive
its transformation by making use of the invariance of the speed of light. We
know how the energy of the photon transforms in {\sc DSR} because the Lorentz-transformations
are well known in momentum space. The speed of the photon is then just the value
of the speed of light $c(E'_\gamma)$ at the value of the energy $E'_\gamma$ that is the result
of the {\sc DSR}-transformation of $E_\gamma$. With that we know the angle of the worldline. To
have the full transformation, we need to fix one point on the worldline. This can most
easily been done by using the notion of the fixed point for this {\sc DSR} transformation, previously
introduced in \cite{Hossenfelder:2010jn}\footnote{Note that the term `fixed point' in the context
of DSR is in \cite{Smolin:2010mx} used differently.}.

Consider making a transformation of the worldline of the particle into another restframe
twice, once by use of the standard special relativistic transformation, and once by use of the
{\sc DSR}-transformation. Since after the transformation the angles of both lines differ,
they will meet (in 1+1 dimensions and flat space) in exactly one point. The
fixed point is this one point on the worldline the transformation of which does not
depend on whether one uses the special relativistic or the {\sc DSR} transformation. This
point does always exist in 1+1 dimensions, and in 3+1 it can always be chosen to exist,
but we will in the following restrict our attention to the 1+1-dimensional case.

Choosing the fixed point then entirely determines the transformation of the worldline of the 
highly energetic photon. In \cite{Hossenfelder:2009mu,Hossenfelder:2010tm} the
fixed point has been in the gamma ray burst. 

Since the {\sc DSR} transformation differs from the usual special relativistic
transformation, in the second restframe the highly energetic photon misses the
electron and only catches up with it well outside the detector. That is the
problematic nonlocality: What is a point-interaction in one frame becomes
a highly nonlocal interaction in another frame. Since all frames are equivalent
we can use well-tested particle interactions to constrain the possibility of
there being such nonlocality. This results in a bound of approximately $\alpha < 10^{-23}$.
As already pointed out in \cite{Hossenfelder:2009mu,Hossenfelder:2010tm}, one should not take the last 
5 orders of magnitude of this bound too seriously because
from $\alpha \approx 10^{-18}$, effects of second order in $E_\gamma/m_{\rm p}$ would become 
non-negligible and those were not taken into account in the derivation of the bound. One can 
however safely conclude that a first order modification is ruled out.

It should be clear from the above 
explanation that if one chooses a different fixed point, then the problem with nonlocality
one obtains is just a translation of the originally considered one. The nonlocality
always becomes macroscopic at large distance (here $\approx$~4 Gpc) from the
fixed point. By definition the fixed point is the point close by which deviations
from special relativity are small and thus problems with nonlocality negligible. 

In summary, we see that two assumptions have been made to arrive at the
scenario considered \cite{Hossenfelder:2009mu,Hossenfelder:2010tm}: That $c(E)$
is observer-independent and that the fixed point of the transformation is
in the gamma ray burst. We also see that only one of these assumptions, the
observer-independence of $c(E)$, is relevant to arrive at the bound. The
location of the fixed point merely determines where the nonlocality becomes
relevant, but it does not change its disastrous consequences for elementary
particle physics.

\section{Smolin's paper}

For his argument, in addition to the observer-invariance of $c(E)$, Smolin needs to 
make the following assumptions
\begin{enumerate}
\item The fixed point of the transformation is always the origin of the coordinate system.
\item Any detector is always at the origin of the coordinate system.
\item Any observer is always at the origin of the coordinate system.
\end{enumerate}

The first assumption is a consequence of the particular framework that Smolin
is using to derive the {\sc DSR}-transformation. This assumption/use of the framework is of course possible,
just that one then needs to keep in mind with this identification one can no
longer independently chose the location of the fixed point and the origin.
Consequently, to reproduce the scenario in  \cite{Hossenfelder:2009mu,Hossenfelder:2010tm},
the origin of the coordinate system would have to be put at the source\footnote{The scenario in \cite{Hossenfelder:2009mu,Hossenfelder:2010tm}
does not depend on the choice of the origin of coordinates, it merely depends on the location of the fixed point.}. This first
assumption is actually unnecessary for Smolin's argument. One can drop it
and replace `origin of the coordinate system' with `fixed point' in the second and third assumption, but
for better comparison to his paper, we will stick with it.

With the first, the second assumption is necessary to prevent nonlocality at
the detector. The third assumption is necessary so the observer is not
sitting in the region where nonlocal effects are sizeable. 
The motivations Smolin puts forward for assumption two and three is that
it is natural to locate the detector at the origin of the coordinate system, 
and that in the usual Einsteinian synchronization procedure one commonly locates the 
observer at the origin. That is indeed an often used and convenient choice,
but in ordinary special relativity it is not necessary, and thus
there is no good reason why this choice in particular should be carried
over to {\sc DSR} where it then gains additional relevance. The assumptions
are in combination not even very realistic: An observer is in most cases 
not in the detector, they can in fact be arbitrarily far away from the
detector, even though it might then take a while to receive the signal. 
Anyway, while Smolin's assumptions seem artificial, we will in the following 
just accept them and see where they take us.

With these assumptions being made, Smolin still has a problem with
nonlocality. There is now no nonlocality at the detector where an observer, D,
sits, but the {\sc DSR} transformation brings up the nonlocality instead 
at the source. By virtue of assumptions 1-3, we are prohibited from
just putting an observer, S, in the same coordinate system as D but
located at the source, where he would sit directly in the middle of the
nonlocality. Instead, all we can do is consider another observer at the source, S', 
that one differing from S by the location of the origin of their coordinate systems. 
Since S' has his fixed point
at the origin and thus at the source, he sees no nonlocality there. This is in conflict with
what D's {\sc DSR} transformation says though. Since the observer S'
at the source sees no nonlocality at the source and S does not exist by assumption 3, Smolin argues that 
the nonlocality at the source that appears in D's transformation is an unphysical coordinate artifact.

So far so good. But at this point one is left to wonder about two
things. 

First, if the nonlocality is a coordinate artifact of the
transformation and the transformation thus only reliable close
by the origin, then what is the physically relevant transformation? 
Basically, Smolin is saying that we can only consider the {\sc DSR}-transformation
to be a local transformation. But then what is the global transformation?
Well, the only global transformation that has
no nonlocality neither at the detector nor at the source is one that has
a fixed point both at the detector and the source, and is thus just the
usual special relativistic transformation. This necessitates together
with the low-energy limit of the speed of light that we are dealing with
an energy-independent speed of light in ordinary special relativity\footnote{
Without using the low-energy limit of the speed of light, one
can still have an energy-dependent speed of light that is compatible with
special relativity, but it corresponds to a massive (possibly tachyonic) photon and
has nothing to do with DSR.}.

Second, Smolin hints at there being just no global transformation,
but only a set of local patches. To begin with, this is unsatisfactory
as an offer for a novel approach to {\sc DSR} since it would be necessary
to know how to describe the propagation of a particle in the full
spacetime, and not just locally, to even be able to tell whether or not
the new approach circumvents the bound while preserving the prediction about
the time delay.
But besides this, it also begs the question why 
not the observer D can just construct the physically relevant global 
transformation from synchronization with the observer S'. The
argument Smolin offers against this is that there is a limit to how accurately
the both observers can synchronize their clocks, respectively compare their
coordinates. If that was true, then it might indeed not be meaningful
to ask for the global transformation and one could not conclude
that one has to get back just special relativity. 

So let us look closer at the synchronization procedure. We note
that Smolin is talking in his paper about a purely classical setup.
In this case, while an energy-dependent speed of light alters the
synchronization procedure, it does not add ambiguity to it. So 
the additional limitations that he is talking about were 
probably meant to be quantum effects (as discussed in Version 2.1. of the box
problem in \cite{Hossenfelder:2009mu}). This does indeed cause an
uncertainty to the sending back and forth of light signals that
becomes more relevant at high energies and is an excellent point to make. 
One might think that to prevent this uncertainty, one could just take the 
low energy limit, get rid of the {\sc DSR}-effects, and recover special relativity. However, in the low energy 
limit the precision of the synchronization procedure is limited by the
usual uncertainty. Taken together, the total uncertainty in the
synchronization procedure with photons of energy $E$ is:
\beqn
\Delta x_{\rm tot} = \frac{1}{E} + \alpha \frac{E}{m_{\rm p}} L \quad.
\eeqn
The most precise synchronization can then be achieved with energy 
\beqn
E_{\rm best} = \sqrt{\frac{m_{\rm p}}{\alpha L}} \quad,
\eeqn
and the corresponding uncertainty is
\beqn
\Delta x_{\rm best} = 2 \sqrt{\frac{\alpha L}{m_{\rm p}}} \quad.
\eeqn
We note that this is equal to $\sqrt{\Delta T/E_\gamma}$. One can now compute the maximal possible precision of the synchronization for the
case of $\alpha = 1$. One finds that the best energy is $\approx 10^{-2}$~eV and the corresponding
precision $\approx 10^{-4}$~m. This has to be contrasted to the nonlocality, which in this
case for a small boost to a relative velocity of $v = 10^{-5}$ already amounts to approx 1 km.
Thus, clearly the observer D can find out that the {\sc DSR}-transformation he has been
using is in conflict with reality, there is nothing preventing him from this insight. 

However, we note that the uncertainty is only proportional to $\sqrt{\alpha}$ while the
nonlocality is proportional to $\alpha$. This means, if we decrease $\alpha$, we will
reach a value where the uncertainty indeed prevents the synchronization procedure
from conflicting with the result of the {\sc DSR}-transformation. In the following, we use
the results from \cite{Hossenfelder:2009mu}, just translated to the source to see what
value of $\alpha$ is necessary to achieve this. The
nonlocality is caused by a mismatch between the transformation behavior of
a worldline under ordinary special relativistic transformation and the modified
{\sc DSR}-transformation. This mismatch has in \cite{Hossenfelder:2009mu} been
denoted as $\Delta T' - t_a'$. The primes indicate that the quantities are those
of a restframe in relative motion to the first (Earth) restframe. The mismatch vanishes for zero relative
velocity, since $\Delta T = t_a$ by construction. (For details and figures, please 
refer to \cite{Hossenfelder:2009mu}.) 

The requirement that Smolin's argument holds and
the observers cannot use the synchronization to discover the {\sc DSR} coordinate artifact then
means the mismatch has to be smaller than the maximally possible resolution in the
second restframe, ie:
\beqn
\left| \Delta T' - t_a' \right| \lesssim \Delta x'_{\rm best} \quad.
\eeqn
With the moderate boost to a relative velocity of $10^{-5}$ used for the original example with the satellite, ie. $v\ll 1$, one gets
\beqn
10^{-5} \Delta T \lesssim \sqrt{\frac{\Delta T'}{E_\gamma'}} \quad. \label{dort}
\eeqn
From the transformation of $\Delta T$ \cite{Hossenfelder:2009mu,AmelinoCamelia:2010qv} 
\beqn
\Delta T' = \frac{1-v}{1+v} \Delta T \quad,
\eeqn
(neglecting terms of second order in $E_\gamma/m_{\rm p}$) and the transformation of $E$ being (also to first order in $E_\gamma/m_{\rm p}$) just the usual
relativistic redshift, one has
\beqn
\frac{\Delta T'}{ E_\gamma'} = \sqrt{\frac{1-v}{1+v}} \frac{\Delta T}{E_\gamma} + {\mbox{higher order}}\label{hier}
\eeqn
We remind the reader that in the original example $v<0$. Since the transformation-factor in Eq. (\ref{hier}) is
for small $v$ approximately 1, we find with $E_\gamma \Delta T \approx \alpha 10^{24}$ from Eq. (\ref{dort}) that
\beqn
\alpha \lesssim 10^{-14} \quad.
\eeqn
That is already quite a tight constraint on Smolin's scenario. But, as argued in \cite{Hossenfelder:2009mu,Hossenfelder:2010tm},
we have tested boosts up to $\gamma = 30$ and found no disagreement with ordinary special relativity. Translation invariance
allows us to apply these at the source as well as the detector, and we will thus have to look at the constraint on
$\alpha$ for $\gamma$ up to 30. Then, one has $\gamma \approx 1/\sqrt{2 \epsilon}$ with
$\epsilon = 1 + v \approx 10^{-3}$ and the relevant uncertainty (cmp to Eq. (18) in \cite{Hossenfelder:2009mu}) is expressed through
\beqn
\left| \frac{2}{\epsilon} - \sqrt{\frac{\epsilon}{2}} \right| \Delta T \lesssim \Delta x'_{\rm best} = \sqrt{\frac{\Delta T'}{E_\gamma'}}\quad.
\eeqn
Since 
\beqn
\frac{\Delta T'}{E_\gamma'} \approx \sqrt{\frac{2}{\epsilon}} \frac{\Delta T}{E_\gamma} \quad,
\eeqn
and one can neglect the term in the inequality proportional to $\epsilon$, one obtains
\beqn
\frac{2}{\epsilon} \lesssim \left( \frac{2}{\epsilon} \right)^{1/4}  \sqrt{\frac{1}{E_\gamma \Delta T}} \quad.
\eeqn
Inserting $\epsilon$ and $E_\gamma \Delta T$ one finds
\beqn
\alpha  \lesssim 10^{-28} \quad.
\eeqn
Which again rules out a first order modifications of {\sc DSR}, even in Smolin's new
interpretation. However, to be very clear here: this bound is of a fundamentally different nature than the
originally derived one. The originally derived bound resulted from constraints on
nonlocal effects that appear in the usual {\sc DSR} scenario. The bound derived
here is a bound necessary for Smolin's scenario to be consistent, such that
nonlocal effects can be interpreted as coordinate artifacts.

The reason for the constraints that we have arrived at here is that we have
studied a synchronization procedure that allows a better precision than the
one considered by Smolin in section 4 in his paper, leading to his Eq. (30). 
The difference is that we have first calculated the energy at which photons
yield the best possible precision for the synchronization. 

It remains to be said that
while we have restricted ourselves here to distances, energies, and boosts that are 
accessible by today's experiments, the conceptual problem is far worse. If the
theory was to apply to all boosts, then the nonlocality would become arbitrarily
large in arbitrary vicinity of all observers and detectors. Or, to put it differently, the
requirement that no two observers under no circumstances are able to make
a synchronization in conflict with the {\sc DSR}-transformation would just result
in $\alpha = 0$, ie special relativity.

\section{Summary}

Smolin suggests to
interpret the nonlocality in deformed special relativity as coordinate
artifacts, arising from applying a local coordinate transform out
of its range of applicability. First, we have summarized which additional 
assumptions are necessary to make this interpretation possible. Then, we have then seen that if 
one takes this interpretation seriously, one must require that no two observers are 
able to use a synchronization procedure whose results are in conflict with 
the (now local) {\sc DSR}-transformation, because otherwise they had a
way to physically construct a better transformation. Since the uncertainty
in the synchronization procedure does not scale the same way as the 
coordinate artifacts, there exist parameter ranges in which Smolin's
scenario is consistent. However, these parameter ranges do again not
allow a modification to first order in energy over Planck mass.

Thus, the statement made in \cite{Hossenfelder:2009mu,Hossenfelder:2010tm} still
holds: {\sc DSR} effects, if they exist, cannot be the source of an observable
time delay in the arrival times of highly energetic photons from distant gamma ray bursts. 
It should be emphazised that this of course does not mean there is no observable signature in
the signal from the gamma ray bursts and that one should not look for it. It just that means that, at the present 
status of discussion, any such signature is
very likely to be of astrophysical rather than of quantum gravitational origin.

\section*{Acknowledgements}

I thank Stefan Scherer and Lee Smolin for helpful discussions.


\begin{thebibliography}{99}


\bibitem{AmelinoCamelia:2000mn}
  G.~Amelino-Camelia,
  {\it ``Relativity in space-times with short-distance structure governed by an
  observer-independent (Planckian) length scale,''}
  Int.\ J.\ Mod.\ Phys.\  D {\bf 11}, 35 (2002)
  [arXiv:gr-qc/0012051].

\bibitem{DSR}   
  J.~Kowalski-Glikman,
  {\it ``Observer independent quantum of mass,''}
  Phys.\ Lett.\  A {\bf 286}, 391 (2001)
  [arXiv:hep-th/0102098];
  J.~Magueijo and L.~Smolin,
  {\it ``Lorentz invariance with an invariant energy scale,''}
  Phys.\ Rev.\ Lett.\  {\bf 88}, 190403 (2002)
  [arXiv:hep-th/0112090];
  J.~Magueijo and L.~Smolin,
  {\it ``Generalized Lorentz invariance with an invariant energy scale,''}
  Phys.\ Rev.\  D {\bf 67}, 044017 (2003)
  [arXiv:gr-qc/0207085].

\bibitem{ReviewGAC}  
  G.~Amelino-Camelia,
  {\it ``Doubly-Special Relativity: Facts, Myths and Some Key Open Issues,''}
  Symmetry {\bf 2}, 230 (2010)
  [arXiv:1003.3942 [gr-qc]];
\bibitem{ReviewJKG}
  J.~Kowalski-Glikman,
  {\it ``Doubly special relativity: Facts and prospects,''}
  arXiv:gr-qc/0603022.



\bibitem{Hossenfelder:2009mu}
  S.~Hossenfelder,
  {\it ``The Box-Problem in Deformed Special Relativity,''}
  arXiv:0912.0090 [gr-qc].

\bibitem{Hossenfelder:2010tm}
  S.~Hossenfelder,
  {\it ``Bounds on an energy-dependent and observer-independent speed of light from
  violations of locality,''}
  Phys.\ Rev.\ Lett.\  {\bf 104}, 140402 (2010)
  [arXiv:1004.0418 [hep-ph]].

 

\bibitem{Smolin:2010xa}
  L.~Smolin,
  {\it ``Classical paradoxes of locality and their possible quantum resolutions in
  deformed special relativity,''}
  arXiv:1004.0664 [gr-qc].

\bibitem{Hossenfelder:2010yp}
  S.~Hossenfelder,
  {\it ``Comments on Nonlocality in Deformed Special Relativity, in reply to
  arXiv:1004.0664 by Lee Smolin and arXiv:1004.0575 by Jacob et al,''}
  arXiv:1005.0535 [gr-qc].

\bibitem{Jacob:2010vr}
  U.~Jacob, F.~Mercati, G.~Amelino-Camelia and T.~Piran,
  {\it ``Modifications to Lorentz invariant dispersion in relatively boosted
  frames,''}
  arXiv:1004.0575 [astro-ph.HE].
 

\bibitem{AmelinoCamelia:2010qv}
  G.~Amelino-Camelia, M.~Matassa, F.~Mercati and G.~Rosati,
  {\it ``Taming nonlocality in theories with deformed Poincare symmetry,''}
  arXiv:1006.2126 [gr-qc].

\bibitem{Hossenfelder:2010jn}
  S.~Hossenfelder,
  {\it ``Reply to arXiv:1006.2126 by Giovanni Amelino-Camelia et al,''}
  arXiv:1006.4587 [gr-qc].

\bibitem{Smolin:2010mx}
  L.~Smolin,
  {\it ``On limitations of the extent of inertial frames in non-commutative relativistic spacetimes,''}
  arXiv:1007.0718 [gr-qc].


 

\bibitem{Ellis:2002in}
  J.~R.~Ellis, N.~E.~Mavromatos, D.~V.~Nanopoulos and A.~S.~Sakharov,
  {\it ``Quantum-gravity analysis of gamma-ray bursts using wavelets,''}
  Astron.\ Astrophys.\  {\bf 402}, 409 (2003)
  [arXiv:astro-ph/0210124].

\bibitem{Jacob:2008bw}
  U.~Jacob and T.~Piran,
  {\it ``Lorentz-violation-induced arrival delays of cosmological particles,''}
  JCAP {\bf 0801}, 031 (2008)
  [arXiv:0712.2170 [astro-ph]].

 
  
  






 
\end{thebibliography}
\end{document}